\documentclass{svjour3}

\usepackage{multirow}
\usepackage{enumitem}
\usepackage{tabularx}
\usepackage{multirow}
\usepackage{graphicx}
\usepackage{colortbl}
\usepackage{booktabs}
\usepackage{natbib}
\usepackage[utf8]{inputenc} 
\usepackage{soul}

\begin{document}

\title{Understanding and Supporting the Design Systems Practice}

\author{Yassine Lamine \and
        Jinghui Cheng
}

\institute{Yassine Lamine \at
                HCDLab, Polytechnique Montr\'eal, Montr\'eal, QC, Canada\\
                \email{yassine.lamine@polymtl.ca}
           \and
           Jinghui Cheng \at
                HCDLab, Polytechnique Montr\'eal, Montr\'eal, QC, Canada\\
                \email{jinghui.cheng@polymtl.ca}
}

\date{Received: date / Accepted: date}

\maketitle

\begin{abstract}
Design systems represent a user interaction design and development approach that is currently of avid interest in the industry. However, little research work has been done to synthesize knowledge related to design systems in order to inform the design of tools to support their creation, maintenance, and usage practices. This paper represents an important step in which we explored the issues that design system projects usually deal with and the perceptions and values of design system project leaders. Through this exploration, we aim to investigate the needs for tools that support the design system approach. We found that the open source communities around design systems focused on discussing issues related to behaviors of user interface components of design systems. At the same time, leaders of design system projects faced considerable challenges when evolving their design systems to make them both capable of capturing stable design knowledge and flexible to the needs of the various concrete products. They valued a bottom-up approach for design system creation and maintenance, in which components are elevated and merged from the evolving products. Our findings synthesize the knowledge and lay foundations for designing techniques and tools aimed at supporting the design system practice and related modern user interaction design and development approaches.
\end{abstract}


\section{Introduction}\label{sec:Introduction}

Design system is a concept that emerged during recent years in the industry that focuses on the design and development of information and communication technologies. They can be understood as ``a collection of reusable components, guided by clear standards, that can be assembled together to build applications''~\citep{dshandbook}. The prominent examples of design systems include Google's Material Design and Microsoft's Fluent Design System. Motivations of adopting a design system are usually originated from the various design challenges that the companies and organizations experience, which are associated with the consistency of the UI design, maintainability of the design work along with its code, and collaboration among the design and development teams~\citep{dshandbook}. Take Google's Material Design as an example, it is a design system that is currently used across multiple Google products (e.g Gmail, Youtube, Google drive, etc). It includes a variety of reusable UI components such as inputs, alerts, navigation, and layout. It has a large open-source community and is used by developers and companies other than Google to build their products after customizing and adapting it for their needs.

Design systems help standardize user interaction design within an organization by creating guidelines and reusable components that can be adopted across multiple products and/or product versions. They ensure that the desired system characteristics concerning usability (such as efficiency, accessibility, and performance) are consistently met while helping designers and developers efficiently build cross-platform user interfaces. Many industrial practitioners have discussed the benefits of this approach and provided information and suggestions for creating design systems~\citep{Fessenden2021, Fanguy2019}. However, in our preliminary conversations with UI/UX practitioners, the adoption of the design systems approach in real-world organizations is often not a smooth process. This is despite the recently available tools such as the InVision Design System Manager\footnote{https://www.invisionapp.com/design-system-manager} and Zeplin\footnote{https://zeplin.io}.

On the other hand, there is very little research work on the design system practice until very recently~\citep{Churchill2019}. The few recent studies on this topic have mostly focused on demonstrating the benefits, values, and impacts of this practice~\citep{Yew2020, Vendramini2021}. Little knowledge has been synthesized across different companies and organizations about the challenges and best practices that can inform a tool design. There is not even a consensus on the definition of design systems. To make the problem more complicated, mature design systems are often surrounded by large communities comprised of members beyond the organizations that own the design systems and the associated projects. The interests and concerns of these broader communities are important factors that affect the evolution of the design system projects, factors that should also be considered in design system tools.

In this paper, we aim to synthesize the knowledge about issues often raised by design system communities and the concerns and perceptions of design system project leaders. This knowledge will be able to provide recommendations and guidelines to inform the design and the evolution of tools to support this new approach. Particularly, we address these aspects by targeting the following research questions:

\textbf{RQ1: What types of issues do the design system projects usually deal with?} Many design systems are surrounded by large communities that have diverse concerns and interests. The design system creators and maintainers need to consider and address issues raised by these communities in order to make their projects successful. Understanding the type of these issues and their frequencies would be able to help assess the major burdens in design system practitioners' work and inform features that can be included in tools for design system creation and maintenance. Thus in this RQ, we aim to create a quantified taxonomy of issues raised by communities surrounding design system projects.

To answer RQ1, we used discussions in issue tracking systems (ITSs) of open-source design systems as a proxy to understand the communities' concerns and the issues to be addressed by design system projects. In our preliminary exploration of design system projects, we found that although usually considered as private property, there is a trend of putting design systems as open source projects. Many of these projects are already attracting large communities beyond the companies and organizations that own the design systems. We conducted a content analysis of the data collected from issue discussions in 41 open-source design systems hosted on GitHub. 
Through analysis of 4714 issues about these design systems, we found that open source design system communities used the ITSs to mainly discuss issues around the user interface components included in design systems; they seemed to be focused more on the behaviors of these components rather than their visual design. Documentation issues were also an interest in the community but were less represented than we expected. The issue tracking systems were also used to discuss community-related topics, such as ways to attract external contributors, as well as software development issues, such as deployment strategies and dependency-related bugs.

\textbf{RQ2: How do leaders of design system projects currently perceive and approach this new practice?} Tools for design system creation and maintenance should be focused on addressing the challenges faced by the practitioners and incorporate the best practices currently adopted in design system projects. Additionally, an accurate definition of the design system should be derived from how practitioners perceive and approach this practice. Thus in this RQ, we aim to synthesize information about the definition, major challenges, and effective strategies around design systems from the perspective of design system project leaders; this sample represent the most influential individuals that initiated and drive the design system approach.

To answer RQ2, we conducted in-depth interviews with nine design system project leaders who played crucial roles in the creation and maintenance of nine different well-known design systems. These participants collectively defined a design system as \textit{an all-in-one design and development environment that includes both UI components and guidelines and serves as a common language for unifying design.} Participants discussed challenges involved in the creation and use of design systems including balancing customizability and standardization, managing their evolution, and addressing unfavorable culture or mindset of the organization. To overcome these challenges, our participants emphasized the importance of a bottom-up process that distills the design systems from existing products and evolves the design systems with those concrete products.

Put together, this study represents an important step towards understanding the practice of creating, maintaining, and using design systems, as well as synthesizing knowledge to inform tools that support this practice. Our results will inspire more research in this direction and pave the road for future tools and methods focused on this and other important design-related practices.
\section{Background and Related Work}
Our work is situated in (1) the literature on design patterns and frameworks, (2) the recent exploration and the gray literature on design systems, (3) studies focused on understanding user interaction designers, (4) previous work on collaboration between designers and developers, and (5) the literature on open source development.

\subsection{Design Patterns and Frameworks}
The concept of design systems has a intimate link to the extensive previous work on design patterns and frameworks. Design patterns originated from Christopher Alexander's work in the field of architecture to describe a collection of common solutions that have solved recurring problems in corresponding design contexts~\citep{Alexander1977Pattern}. It has been then adopted in many fields, including user interaction design. For example, \cite{Erickson2000} conceptualized the interaction design pattern language as a \textit{lingua franca}, a common language that supports communication among various kinds of stakeholders (including users) in a design process. \cite{Tidwell2020} has summarized more than 100 interactive design patterns, put into 11 categories; this pattern library has covered a wide range of aspects in user interaction design, such as content organization, navigation, form design, and data visualization. Many researchers have also focused on identifying design patterns in a specific domain, including web apps~\citep{Duyne2002,Scott2009}, mobile apps~\citep{Neil2014}, information retrieval systems~\citep{Morville2010}, ubiquitous computing~\citep{Chung2004}, video games~\citep{Cheng2017CHIplay,Bjork2004}, and, most recently, intelligent systems~\citep{Gutzwiller2018,Ma2019}. In general, user interaction design patterns suggest high-level design solutions based on specific problems that the designers face. They are usually descriptive and include elements such as a name (for shortcut communication), a problem statement, a solution description, and examples that realized the solution. Research has identified that design patterns are useful tools in participatory design and stakeholder communication~\citep{Cheng2017CHIplay}; however, communication breakdowns can still appear when design pattern is an unfamiliar concept~\citep{Dearden2002}.

Different from design patterns, front-end development frameworks focused on supporting the implementation of user interface elements~\citep{Saxena2019}. Popular frameworks include Bootstrap\footnote{https://getbootstrap.com}, Foundation\footnote{https://foundation.zurb.com}, Pure\footnote{https://purecss.io/}, and Siimple\footnote{https://www.siimple.xyz/}. They usually include reusable and customizable code to support the construction of UI layout (e.g. many include a grid system for easy layout) and the creation of UI components (e.g. buttons, lists, navigation bars, etc.). The development of these frameworks is often managed by an organization or company. However, they are aimed to serve the purpose of general UI design and implementation.

While design patterns and frameworks are related concepts, design systems focused on different aspects of design support. Different from patterns, which usually discuss high-level design solutions, design systems incorporate specific guidelines and components to create concrete support for both design and implementation. For example, the Google Material design system does not only provide a set of design guidelines and principles, but it also includes the code for concrete UI components or component groups for fast prototyping and development. Different from frameworks, which often aimed to support general UI design and implementation, design systems are more organization-specific, incorporating branding-related elements and artifacts. For example, in the Shopify Polaris design system, the Shopify brand is built into various elements from color schemes, component styles, and workflows that the components enable, in order to support a coherent user experience on the various apps of the platform. Our study builds on top of the work on the related areas of design patterns and frameworks to explore design systems as a less represented but highly impactful topic.

\subsection{Design Systems}
The concept and practice of design systems have recently gained considerable attention in industrial settings. Practitioners have shared experiences and knowledge of creating, managing, and using design systems in websites and blog posts. We briefly review these documents as gray literature. For example, \cite{kholmatova2017design} compared design systems with design patterns and advocated design systems for going beyond patterns to provide ``techniques and practices for creating, capturing, sharing and evolving those patterns.'' The author recognized that there was not a commonly accepted definition of ``design systems'' and recommended a process for creating design systems based on a conceptual separation of functional patterns and perceptual, or visual, patterns. \cite{Hacq2018} emphasized that the design systems contain a set of deliverables that include a style guide that describes ``graphic styles and their usage'' and a pattern library that integrates the functional components. \cite{Fanguy2019} synthesized the resources about building a design system and recommended a four-step process that includes: (1) a visual audit of the current design, (2) design language creation, (3) UI pattern library creation, and (4) documentation. \cite{Fessenden2021} also summarized the principles of the design systems approach and defined a design system as ``a complete set of standards intended to manage design at scale using reusable components and patterns.'' They also recognized that design systems contain two important elements: (1) the design repository that includes all the related artifacts and (2) the team that manages the design system. There were also yearly efforts to conduct surveys with the design system community \citep{Yew2019, Hamilton2020} to understand the current trend in the practice and tooling.

Along with the practical interests, research on the topic of design systems emerged during the past a few years, initially done by researchers in industry. \cite{Churchill2019} wrote a column article to reflect on the design system practice at Google and described it as having value for reducing effort, scaffolding learning, increasing collaboration, and aligning process and product. Based on these values, the author called for more research work on this topic. Google researchers \citep{Yew2020} also conducted a survey study with the design systems community and found that UI consistency and brand are factors that contributed to both the motivations and the values of the design systems. Beyond the traditional graphical user interfaces, researchers from IBM \citep{Moore2020} also explored design systems in the context of conversational UI.

It is only until very recently that academic studies started to discuss the design system practice. We only found two peer-reviewed studies in this area that we summarize below. \cite{Vendramini2021} conducted a review of both white and gray literature and identified 23 sources discussing the design system practice. The authors identified that this literature usually described design systems as including three key elements: component libraries, design guidelines, and style guides. They also summarized a set of benefits and impacts of adopting this practice. \cite{Handal2022} described a case study, at a very high level, in which a large company adopted the design system approach. The authors briefly described the adoption process and the impacts after adoption.

Despite these industrial and academic research efforts, there is still a lack of consensus on the concept of design systems and taxonomy to characterize community concerns. There is also limited investigation of challenges and strategies related to the creation and maintenance of design systems. Our study aims to fill these gaps.

\subsection{Understanding User Interaction Designers}

User interaction designers are a special group of practitioners who links the users' needs and desires to the technical abilities provided in software systems. On one hand, they share many characteristics of all professional practitioners, considering, as Sch\"on identified, the specific practical problem through ``reflection-in-action'' (i.e., constant assessing and adjusting actions in an unfolding situation)~\citep{schon1984reflective}. On the other hand, UI designers tackle the unique design task, which Cross argued to be ``ill-defined'' and requires solution-focused strategies~\citep{Cross2001}. In a seminal paper, \cite{Gould1985} outlined three principles that defined a ``user-centered'' approach: (a) early focus on the user, (b) empirical measurement, and (c) iterative design. Since then, much work has been put to understand user interaction designers and their design practices. For example, \cite{Stolterman2012} advocated the concept of \textit{Designerly Tools} aimed at exploring ``methods, tools, techniques, and approaches that support design activity in a way that is appreciated by practicing designers''. They have found that the designers used various tools to support two different types of activities: (1) design thinking and ideation and (2) the creation of design artifacts. Leveraging these concepts, \cite{Gray2016} has identified that interaction designers rely on a user-centered mindset to guide their adoption of design methods and tools. \cite{Zhang2014} have also pointed out that the personal experience of interaction designers can have a major impact on their practice.

The methods adopted in our study were informed by the insights gained from this previous literature. Particularly, we consider design systems a type of ``designerly tools'' to support both (1) ideating consistent and high-quality design through consideration of the design guidelines and (2) efficiently constructing mockups and working user interfaces with reusable components. We investigate how designers and the broader communities work with such a tool.

\subsection{Collaboration Between Designers and Developers}

Collaboration between designers and developers is an essential aspect of the development of user-facing software. \cite{4293575} investigated the integration of UI/UX design into agile development. They reported on a case in which embracing the iterative development process of agile had in fact led to the improvement in the developer-designer relationship. \cite{10.1145/2399016.2399121} have conducted a grounded theory field study with software teams and found that collaborative events between designers and developers happen frequently and often go beyond planned activities. Their results also revealed that the majority of the designer-developer collaborations ``implicitly aligned''; in other words, ``designers and developers share an implicit understanding of how collaborative work should be carried out'' and engage in collaborative activities that are not directly targeting work alignment.

While important, the multi-disciplinary nature of the designer-developer collaboration in software organizations can be challenging. For example, \cite{10.1007/11774129_15} described the existence of ``a culture of defensiveness'' issue between designers and developers that can result in communication barriers; based on characteristics of such issue, they proposed a framework to alleviate the tension when integrating development and design. Furthermore, \cite{10.1145/2783446.2783563} investigated designers' perception of the developers' empathy towards designers and design work; they identified that the developers' misunderstandings of the nature of design work have resulted in miscommunications and impacted the quality of the end product. \cite{10.1145/3310276} also found that breakdown during the designer-developer collaboration tended to occur when (1) a specific design detail is not communicated by designers, (2) a particular case is not covered in the design, and/or (3) the design failed to consider developers' technical constraints.

Our work builds on these previous studies to explore the way design systems serve as a medium for facilitating the collaboration between designers and developers. We also explore the challenges design system practitioners face while collaboratively creating and maintaining design systems.

\subsection{Open Source Development}

Interestingly, although usually considered as private property, many design systems have become open source projects that are accessible by other designers and the general public. Open source is a software development model in which the source code of the software product is open for access under a certain license. This development model has gained popularity over the past decades and becomes a common practice in many software-intensive application domains~\citep{Crowston2008,Schrape2019}. Open source software projects are usually hosted on a public repository management platform (e.g. GitHub) and rely on various tools for tasks such as version control, project management, community engagement, and communication. Through these tools, geographically dispersed community members make diverse contributions to the project~\citep{Cheng2019Chase}.

Previous works have suggested that individuals were motivated to join an open source project because of both internal factors (such as socialization~\citep{Gharehyazie2015,Casalnuovo2015}, learning opportunities~\citep{Ye2003}, and self-perceived identity with respect to the project~\citep{Hertel2003}) and external factors (such as human capital and monetary rewards~\citep{Hars2002}). Many researchers and practitioners have pointed out that the development of open source software relies on a healthy community. A traditional view of a typical open source community resembles an ``onion model''~\citep{Nakakoji2002,Ye2003}. This model suggests a hierarchical structure of responsibilities among community members that included a small number of core members and an increasingly larger number of various types of peripheral developers~\citep{Nakakoji2002}. However, recent studies suggested that this structure is not stable and constantly evolving. While the size of the community increases, the boundaries of the hierarchical layers among the peripheral members tend to be blurred~\citep{Joblin2017a}. Additionally, several factors, such as the developer's motivation in participating in different projects and the social structure around the developer, can influence their evolution from a peripheral member to a core member~\citep{Cheng2017}.

One important tool that open source projects often use to engage the community and manage the various tasks is the issue tracking systems (ITSs)~\citep{Bertram2010}. They often provide a forum-like functionality for posting and commenting on issues. Research has established that ITSs contain rich information about the software project~\citep{arya2019analysis} that may include new feature requests, enhancements, bug reports, general tasks to be completed, questions about the software project or product, or even posts that solicit feedback on rough ideas. ITSs play an important role in supporting various software engineering activities such as requirements elicitation and management~\citep{Heck2017,Merten2016}, task and job distribution~\citep{Xia2017}, and traceability management~\citep{Huang2017}, to just name a few. Open course software communities have also often used the ITSs to manage usability-related topics~\citep{cheng2018open,Iivari2011}. In recent work, \cite{Wang2020} have investigated an argumentation model to help the open source community members effectively understand and consolidate usability-related opinions posted on ITSs.

Our project builds on these previous studies and uses issue tracking systems as a proxy to investigate the issues that need to be addressed by design system projects.
\section{Characterizing the Issues to be Addressed by Design System Projects}
To explore the interests and concerns of design system communities and create a quantified taxonomy of issues that need to be addressed in design system projects (i.e. answer RQ1), we used the ITSs of open-source design systems as a proxy and conducted a content analysis on 4714 issues collected from a wide range of open-source design system projects. These issues reflect the focal points of the broad open source communities towards the design systems and represent factors that need to be addressed by the companies and organizations that own the design system projects.

\subsection{Methods}

\subsubsection{Design system project selection}
To identify a wide range of open source design systems, we referred to the Adele repository\footnote{http://adele.uxpin.com}, a curated list of publicly available design systems, style guides, and pattern libraries. We acknowledge that the definition of design systems is not fully established. However, a clear characteristic that differentiates design systems from style guides and pattern libraries is that design systems usually \textit{incorporate the identity of a company or an organization}. We compared the design systems' visual design with the products of their owners using such criteria and selected 52 design systems among the 90 projects listed in the Adele repository. We then only included design system projects that have a public repository on GitHub; we focused on GitHub because of its increasing popularity over the past years among the open source communities as an integrated platform for open source project hosting and management. Finally, we excluded projects that have less than 10 issues and 10 commits. This strategy allowed us to focus on active design system projects and resulted in 41 projects in our dataset. On average, the analyzed projects had 403.0 issues and 42.3 contributors. Table~\ref{fig:design_system_list} summarizes the selected projects.

\begin{table}[ht]
\centering
\resizebox{\columnwidth}{!}{\begin{tabular}{lcccccc}
    \hline
    \textbf{Design system repository} & \textbf{Open issues} & \textbf{Closed issues} & \textbf{Total issues} & \textbf{Contributors} & \textbf{Stars} & \textbf{Forks} \\ \hline
    alfa-laboratory/arui-feather & 12 & 41 & 53 & 64 & 411 & 83 \\
    alphagov/govuk-design-system & 52 & 199 & 251 & 46 & 74 & 65 \\
    Altinn/DesignSystem & 5 & 8 & 13 & 28 & 31 & 11 \\
    audi/audi-ui & 15 & 3 & 18 & 5 & 155 & 25 \\
    auth0/cosmos & 144 & 500 & 644 & 21 & 395 & 85 \\
    auth0/styleguide & 12 & 46 & 58 & 20 & 143 & 51 \\
    brainly/style-guide & 52 & 551 & 603 & 21 & 114 & 17 \\
    bring/hedwig & 13 & 95 & 108 & 21 & 21 & 1 \\
    buzzfeed/solid & 2 & 245 & 247 & 27 & 108 & 20 \\
    cfpb/capital-framework & 64 & 255 & 319 & 17 & 50 & 35 \\
    Dropbox & 14 & 14 & 28 & 12 & 794 & 46 \\
    Financial-Times/ft-origami & 4 & 306 & 310 & 31 & 81 & 12 \\
    FirefoxUX/photon & 58 & 217 & 275 & 31 & 186 & 49 \\
    fs-webdev/fs-styles & 2 & 49 & 51 & 17 & 34 & 17 \\
    gctools-outilsgc/aurora-website & 43 & 87 & 130 & 11 & 13 & 6 \\
    govau/design-system-components & 43 & 307 & 350 & 20 & 584 & 59 \\
    instacart/Snacks & 18 & 34 & 52 & 19 & 50 & 35 \\
    JetBrains/ring-ui & 2 & 874 & 876 & 31 & 2216 & 107 \\
    liferay/lexicon-site & 5 & 22 & 27 & 15 & 21 & 25 \\
    lonelyplanet/rizzo & 6 & 55 & 61 & 57 & 728 & 86 \\
    mesosphere/cnvs & 20 & 19 & 39 & 12 & 27 & 3 \\
    mineral-ui/mineral-ui & 72 & 315 & 387 & 13 & 424 & 42 \\
    mozilla/protocol & 70 & 160 & 230 & 11 & 50 & 26 \\
    OfficeDev/office-ui-fabric-react & 404 & 3294 & 3698 & 388 & 4625 & 1016 \\
    pinterest/gestalt & 22 & 75 & 97 & 44 & 3093 & 181 \\
    pivotal-cf/pivotal-ui & 2 & 261 & 263 & 72 & 605 & 82 \\
    pluralsight/design-system & 43 & 285 & 328 & 20 & 115 & 23 \\
    pricelinelabs/design-system & 58 & 148 & 206 & 34 & 399 & 76 \\
    primer/css & 75 & 195 & 270 & 67 & 8215 & 604 \\
    rei/rei-cedar & 4 & 25 & 29 & 20 & 40 & 11 \\
    salesforce-ux/design-system & 19 & 521 & 540 & 48 & 2530 & 560 \\
    seek-oss/seek-style-guide & 2 & 33 & 35 & 42 & 256 & 37 \\
    Shopify/polaris-react & 184 & 691 & 875 & 126 & 2433 & 385 \\
    SpareBank1/designsystem & 55 & 189 & 244 & 47 & 71 & 40 \\
    sparkdesignsystem/spark-design-system & 197 & 448 & 645 & 17 & 40 & 25 \\
    USAJOBS/design-system & 34 & 119 & 153 & 8 & 27 & 15 \\
    uswds/uswds & 33 & 1480 & 1513 & 98 & 5016 & 691 \\
    uswitch/ustyle & 22 & 113 & 135 & 31 & 16 & 3 \\
    vmware/clarity & 309 & 1820 & 2129 & 51 & 4743 & 458 \\
    vtex/styleguide & 50 & 123 & 173 & 47 & 45 & 7 \\
    wework/plasma & 36 & 23 & 59 & 23 & 17 & 4 \\ \hline
    \textbf{Averages} & 55.5 & 347.4 & 403.0 & 42.3 & 951.1 & 125.0 \\ \hline
    \end{tabular}}
\caption{List of the studied design systems (all counts were conducted in June 2019)}
\label{fig:design_system_list}
\end{table}

\subsubsection{Data collection}
The data collection was conducted in June 2019. We used the GitHub REST API\footnote{https://developer.github.com/v3/} to collect the issue data from our list of design system repositories. Particularly, for each issue, we extracted the title, the description, the comments, the state (open or closed), and other identifying information such as the creator and timestamp. We then selected a random set of 5000 issues from the 16635 issues we extracted from the 41 design system repositories. In a preliminary analysis of these issues, 286 issues were removed due to insufficient or incomprehensible information, resulting in 4714 issues in our final dataset.

\subsubsection{Analysis method}
We conducted a qualitative content analysis on the collected data~\citep{Vaismoradi2013, codfda}. We started our analysis by conducting an inductive coding~\citep{saldana_coding} on a random sample of 200 issues from the collected data to identify the common themes. We particularly focused on (1) the aspects of the design system projects that the issues address and (2) the nature of the issues themselves. This step was first done independently by two researchers and was then followed by a thorough discussion and codes comparison to establish the themes. We then used these themes to code the remaining 4514 issue reports. 

\subsection{Results: Project Aspects}
The communities around open source design systems have raised issues for various aspects of the projects. We categorized the project aspects into the following five prominent groups.

\subsubsection{Behavior of UI components (N~=~2015)} Many issues were focused on the behavior and functionalities of the UI components. Some issues mentioned more than one behavior aspect. The common UI behavior aspects that the design system communities focused on included:
\begin{itemize}[leftmargin=16pt]
    \item \textit{State change behaviors} (245 issues) of UI components to provide appropriate user feedback, for example, when the component is hovered, focused, or disabled.
    \item \textit{UI animation behavior} (153 issues) for improving user experience and attractiveness.
    \item \textit{Accessibility} (124 issues) that elevates the support to people with disabilities to the design system level; many issues are focused on supporting users with visual impairments, considering screen readers and keyboard shortcuts.
    \item \textit{Navigational behaviors} (101 issues) such as pagination and scrolling.
    \item \textit{Input verification} (112 issues) in forms, such as password format or email address verification.
    \item \textit{Keyboard shortcuts} (93 issues) that could enable more efficient user interaction. 
    \item \textit{Responsiveness} (47 issues) that allows UI components and the page structure to adjust to the screen size. 
\end{itemize}

\subsubsection{Visual design of UI component (N~=~480)} 
Some issues discussed by the design systems communities also addressed the visual design of UI components including color scheme, spacing, and typography.

\subsubsection{Documentation (N~=~321)} The documentations are essential in any design system; they do not only describe how the design system should be used, but also discuss important principles and guidelines behind the design system. In most issues about this aspect, contributors usually reported missing documentation, pointed out errors or inaccuracies, and requested to improve documentation of a certain component. For example, in Issue 1647 of the OfficeDev/office-ui-fabric-react project, the issue creator reports an error in the design system version documentation: ``\textit{On the website's home page, we show the version of Fabric Core and Fabric React that the website documents. These are showing as a version range when we want to show the latest version only.''}

\subsubsection{Software development aspects (N~=~310)} Many issues were also focused on the aspects related to the software development process, tooling, infrastructure, programming, and the use of frameworks. The most frequently discussed topics in this category included (1) testing issues (e.g in Issue 183, cfpb/capital-framework, ``The node-wcag tests seem to be failing for many (if not all) cf components...''), (2) deployment and release issues (e.g in Issue 624, mineral-ui/mineral-ui, ``improve the release script to make it easier to use...''), (3) issues related to the design system's code (e.g in issue 4478, OfficeDev/office-ui-fabric-react, ``Component Classes should not define their methods with lambdas...''), and (4) dependency related issues (e.g issue 569, Shopify/polaris-react, ``Move `app-bridge' to peer dependency.''). Additionally, some contributors also discussed the problems of relying on third-party libraries and proposed to reduce such reliance.

\subsubsection{Community (N~=~135)} The issue tracking systems are also used as a communication tool for addressing community-level tasks and processes. Many issues are aimed at discussing ways for attracting external \textbf{contributions} to the design system repository. For example in Issue 725 of the govuk-design-system repository, a community member reported that \textit{``We want to enable more people to contribute, and to make it easier to make smaller contributions to the design system''}. Some issues discussed tasks to help better \textbf{communicate} with the design system users. For example, the reporter of Issue 361 of the auth0/cosmos project suggested a \textit{e.g ``need to publish a changelog (as a documentation) to inform users of changes that we make...''} Issue discussions also revolved around how to better \textbf{satisfy} the needs of the design system users. For example, the reporter of Issue 223 of govau/design-system-components indicated that \textit{``the team needs a method for tracking what technology our users have access to and currently use.''}. Some have also discussed the process and practical issues to support the \textbf{growth} of the community. For example, the reporter of Issue 2301 of the uswds/uswds repository initiated the discussion about the importance to \textit{``assure long-term growth with a small core team.''}

\subsection{Results: Issue Nature}
\begin{figure}[ht]
    \centering
    \includegraphics[width=\columnwidth]{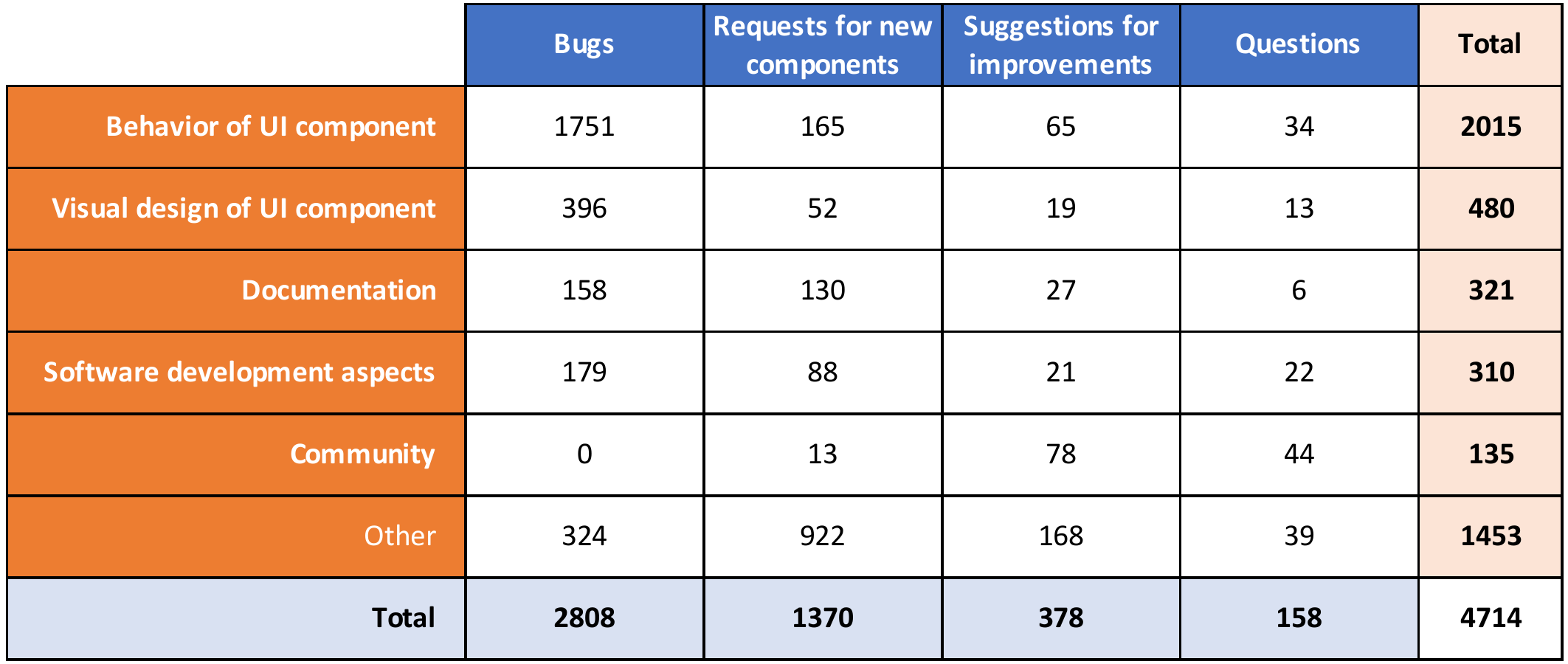}
    \caption{Frequency of project aspects according to issue nature categories}
    \label{fig:issue_frequency}
\end{figure}

In this dimension, we aimed to identify the categories of why the issue is reported.  Figure~\ref{fig:issue_frequency} shows the intersection of these categories with the project aspect groups.

\subsubsection{Bugs or problems in the existing components (N~=~2808)}
About half of the issues we analyzed are reported because of a bug or problem in the existing components (i.e. parts of the existing system that do not appear or behave as expected). While the majority of the bugs or problems were related to UI components, the rest were associated with documentation and peripheral code related to the software development aspects (e.g. deployment code, test code, etc.). These issues usually described the steps or the process to reproduce the bug. For example, Issue 4936 of the OfficeDev/office-ui-fabric-react project reported a color scheme issue that \textit{``Close button for the Message Bar is not visible properly in the High Contrast Black mode.''} 

\subsubsection{Requests for new components (N~=~1370)}
About one-third of the design system issues are also requests for adding new components. The frequently requested components included: (1) \textbf{content containers} such as panel, card, and dialogue; (2) \textbf{navigational components} such as sidebar, breadcrumb, and link; (3) \textbf{informational components} such as banner and progress indicator; and (4) \textbf{page components} such as header, footer, and 404 error content. These requests focused on both unique, complex components (e.g. dialogue, avatar system) and basic, universal ones (e.g. banner, header). This phenomenon indicated that the design system projects are not created as a static library, but are constantly evolving. It also shows that the components in the design systems are carefully selected, may only include the most relevant ones, and are enriched while the projects around it evolve.

\subsubsection{Suggestions for improvements (N~=~378)}
Some issues have also focused on requesting or suggesting improvements to an existing component, the overall functionality of the design system, or in aspects of community engagement and organization. For example, in brainly/styleguide, Issue 365, a contributor is proposing to change the implementation of a certain component: \textit{`` I'd propose to use content-box as a mixin, could be used in places where we use it already, but would require us to create less generic components (i.e mint-ranking, mint-panel etc)).''}.

\subsubsection{Questions (N~=~158)}
Some of the issues posted in design system projects did not report a problem of the system, but instead focused on asking questions and requesting information about design system components, functionalities, and the development process. For example, in Issue 154 of the cfpb/capital-framework repository, the contributor wondered whether they should stop using the auto-generated docs.
\section{Understanding the Perspectives of Design System Project Leaders}

The issue analysis study allowed us to highlight the types of issues open source communities raise about design systems. In order to understand the values and practices of the core contributors of design systems (i.e. to answer RQ2), we conducted an interview study with nine highly experienced design system practitioners who are leaders of their design system projects.

\subsection{Methods}
We conducted the interviews from August to October 2019. In this section, we describe our participants, the interview procedure, and the data analysis approach.

\subsubsection{Participants}
We aimed to recruit practitioners who are highly experienced with design systems creation. In order to identify the qualified participants, we first identified the most active contributors in each of the design system projects we analyzed in our issue analysis study. We defined ``contribution'' broadly to include both committing contributions and issue discussion contributions. Because design system repositories often manage both the code of the component libraries and the documents related to the design pattern and guidelines, major contributors committing to design system repositories would include both developers and designers. Additionally, major contributors to the issue discussions can play the role of core developer, lead designer, and/or project manager, depending on the project. Thus we adopted this recruitment strategy in order to capture the concerns of a wide range of roles that lead design system projects. We contacted the contributors who included a public email address on their GitHub profile in descending order of their contribution. Recruitment was terminated at nine participants based on data saturation~\citep{Faulkner2017Satuation}. Our participants are from nine different companies that own a design system (e.g., Shopify Polaris, Dropbox Scooter, Financial Times Origami, etc.); all have occupied high-level positions related to the design, development, and management of design system projects. In other words, they're directly involved with the most important decisions related to their design systems. All participants are male. Table~\ref{fig:interview_participants_info} summarizes the characteristics of our participants.

\begin{table}[ht]
    \centering
    \begin{tabular}{clcl}
    \toprule
    ID & Country & Design system experience (years) &  Job Title \\
    \midrule
    P1 & UK & 7 & Business owner/developer \\
    P2 & US & 5 & Design lead \\
    P3 & Russia & 3 & Senior product designer \\
    P4 & Spain & 3 & Product design manager \\
    P5 & UK & 6 & Head of engineering \\
    P6 & Canada & 4 & UX development manager \\
    P7 & Spain & 2 & Software engineer \\
    P8 & Spain & 4 & UX design lead \\
    P9 & Canada & 4 & Design system manager \\
    \bottomrule
    \end{tabular}
    \caption{Interview participants information}
    \label{fig:interview_participants_info}
\end{table}

\subsubsection{Interview procedure}
We conducted semi-structural remote interviews using Zoom. Interviews were recorded and each took about 40 minutes to complete. During the interview sessions, we focused on the participants' knowledge and experience in developing or contributing to design systems. Particularly, we started by asking about their definition of a design system considering the fact that the concept is relatively new and unclear. We followed by questions about the benefits of design systems in order to explain their increasing popularity. Finally, we asked about the challenges the participants experienced during the development, maintenance, and usage of design systems; we also asked about the best practices our participants follow to mitigate the challenges.

\subsubsection{Analysis}
To analyze our interview data, we fully transcribed the recordings and performed an inductive thematic analysis~\citep{Vaismoradi2013}. The coding was focused on the following four topics that reflected the main categories of interview questions we asked the participants: (1) concepts used for defining design systems, (2) benefits of design systems, (3) challenges in creating, maintaining and using design systems, and (4) strategies and best practices for mitigating the challenges. Once the inductive coding process is concluded, we created a codebook for describing the themes identified in the categories. Then another researcher (i.e. the blind coder) is involved to use the codebook to code the interview transcripts. We used Cohen's kappa~\citep{cohenkappa} to evaluate the inter-rater reliability between the codebook creator and the blind coder for the codes generated. Among all codes included in the codebook, the average kappa is 0.88, indicating almost perfect agreements between the coders about the identified themes.

\subsection{Results}
\begin{figure}[ht]
    \centering
    \includegraphics[width=\columnwidth]{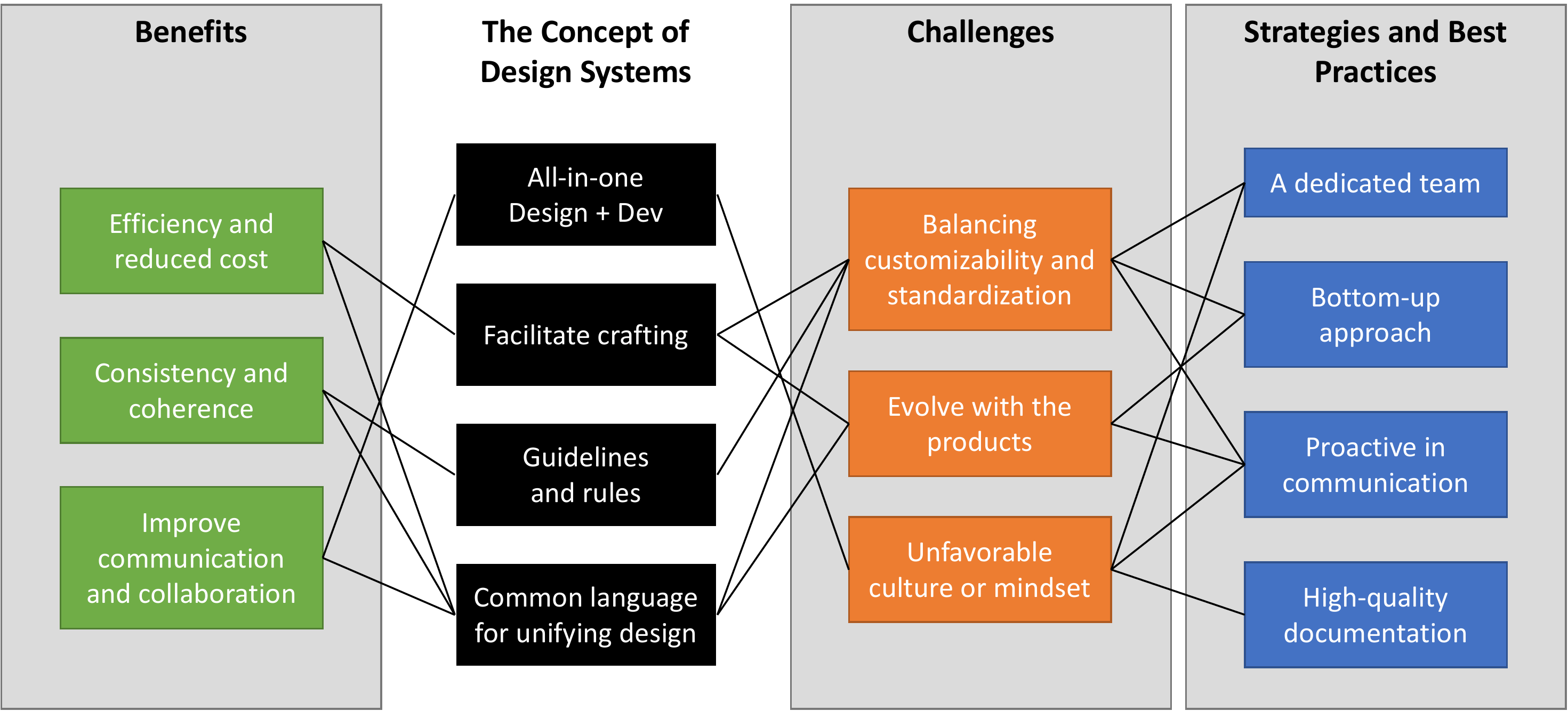}
    \caption{Overview of themes identified in the interviews with design system project leaders.}
    \label{fig:coding-summary}
\end{figure}

In this section, we report the themes identified regarding the participants' discussions around concepts, benefits, challenges, and best practices in the design system approach. Figure~\ref{fig:coding-summary} provides an overview of our results representing the themes in those four categories and the relationships among them.

\subsubsection{Concepts}
Participants described various concepts when asked to define design systems. We grouped these concepts into the following categories.

Four participants described the design system approach as an \textbf{all-in-one design and development environment}. In their opinion, design systems provide all the necessary tools and resources with which all the front-end design and development activities are carried out. As P5 described: \textit{``A design system includes more than just a simple component library or a color palette. It would include the brand guidelines. It also includes fully interactive and shareable components that you can reuse across applications and the web, and whatever other pieces of technology you choose to use there.''}

From the prototyping and development perspective, three participants also considered design systems as a tool for \textbf{facilitating crafting} of prototypes and digital products. Essentially, a design system provides the required components and artifacts for developers to implement design ideas and for designers to quickly produce prototypes. It widely eases their work and serves as a mediating layer for them to focus on more important and/or demanding tasks. For example, P8 mentioned: \textit{``it's the system that includes everything that the designers and the developers need to craft their products''}. P7 also described the design systems as \textit{``a set of components that help us build our product more easily and that help us not having to focus on design and just having to focus on structure. Is more like a helper tool for us.''}

Meanwhile, three participants considered the design system as \textbf{guidelines and rules} for creating UI components. In their minds, the guidelines and rules embedded in the design systems are more important than the reusable UI components. These elements were considered to be a mind-set changing tool to standardize the product design process. They do not only allow multiple teams to work together consistently, but also promote the best practices, according to the needs of the company, in the design and development process. For example, P4 mentioned:  \textit{``... the design system adds another layer on top of the implementation. ... It tells you like, `These are the best practices. These are the limitations that we want you to know. These are the rules.' ''} These practices in turn result in the consistency and coherence of UI components across all the products the company owns, which facilitates the integration of the brand and reinforcement of the identity in the products.

A \textbf{common language for unifying design} is another key concept, mentioned by three participants. They described design systems as tools for communicating the system functionalities to users in a consistent way, using the same components across multiple products, as described by P4: \textit{``basically design system is a language, that allows your product to talk in the same way''}. 
As a unified design language, design systems communicate the UI functionalities consistently to users, helping users to form mental models of the UI components used. This was highlighted by P1: \textit{``I would say a design system is something that allows a user to get the information more easily. ... For example, if we have buttons that are inconsistently colored then a user goes and uses a feature of one of our websites and presses the button that's blue, and then they start to have a mental model of what blue buttons do...''}

\subsubsection{Benefits}
Our participants have discussed various benefits that design systems offer to their companies and teams. We categorized them into the following three high-level themes.

First, seven participants commented on how the use of a design system results in \textbf{efficiency and reduced cost}. With a consistent and centralized design and implementation environment, both developers and designers can follow predefined patterns to accomplish their tasks more efficiently. For example, P2 mentioned: \textit{``You can very quickly go from a nascent idea to a full-fledged prototype or product using a designed system in a way that would be much more difficult without one.''}. P4 highlighted this aspect from the perspective of how the centralized approach of the design system reduces development effort and time. : \textit{``There is a reduction in the effort since there is a centralized point of reference that everyone can follow.''}. 
P3 also stated: \textit{``It's a great economy when we make design and when we develop a product -- we don't have to reinvent the wheel every time.''}

\textbf{Consistency and coherence} is another main benefit of design systems, according to five participants. According to them, using a design system ensures that all the components have a consistent design and implementation across multiple products. With the support of design systems, the UI follows a clear pattern that is centered around the company's brand, which results in consistent products with similarly looking and behaving components. For example, P5 mentioned: \textit{``I think the biggest benefits from a design system are re-usability and consistency. ... For us, the biggest benefits were that shared consistency and understanding of how to best build the components.''}.


Finally, four participants considered that design systems \textbf{improve communication and collaboration} inside IT organizations. They provide a common language not only for the products, but also for the different teams involved in the project development, which results in better collaboration inside the organization. For example, P8 discussed the power of design system as a common language: 
\textit{``It's easier for a designer to communicate something to a developer, like, I need this button and this is a primary button and both are thinking the same thing.''} Participants also valued the design system as a tool for optimizing the onboarding process. New hired developers or designers can easily understand how UI design activities are carried out inside the company. For example, P3 mentioned that, with the support of the design system, \textit{``when the team changes and it happens every time, newcomers can easily join the work and understand how it works.''}

\subsubsection{Challenges}
Despite the benefits, the design systems' centralized approach to creating and maintaining UI design components, patterns, and guidelines posed challenges to practitioners. We summarized three high-level challenges that our participants mentioned.

Ideally, design systems should both reinforce standardized guidelines and components and at the same time allow customization of components based on the needs of individual projects or teams. In reality, however, \textbf{balancing customizability and standardization} in design systems is a challenging task, mentioned by six participants. The simple question about which components to be included in the design system and which to be customizable in individual projects can be complicated. For example, P2, responsible for their design system project, mentioned: \textit{``Sometimes those teams will tell us that they need a new component or pattern in the design system... So, we have to have this debate about whether it is suitable for us to work on that thing, or can we say no and that can mean in a year's time, the team may decide to build something themselves. That's always challenging.''}. 

Related, making sure that the design systems appropriately \textbf{evolve with the products} they facilitate is also difficult, mentioned by six participants. Maintaining consistency and relevance, as well as preventing scope creeping is a tricky process that proves to be much more challenging than the creation of the design system. For example, P6 clarified: \textit{``Developing it, it's not too complicated. It's getting adoption, maintaining it, making sure it's still relevant even after a few years, making sure it stays nimble so you can still update it easily, and it doesn't become crystallized...''}

While design systems can improve designer-developer communication and collaboration, five participants mentioned that the initial \textbf{unfavorable culture or mindset} about communication and collaboration can be an obstacle for the creation, maintenance, and adoption of design systems. For example, P3 discussed: \textit{``If you don't have the right mindset at both designers and developers, ... if the designer turns his design and then developer takes it and they don't talk before, there's a huge chance that you can't create a design system.''} Indeed, the design system approach poses major changes in the way people used to work and can be perceived negatively, as described by P4: \textit{``I get complaints because people used to think that the design system was limiting them somehow. it took a huge effort on communication to tell people that, `Hey, you can do whatever you want. Here is the design system that offers you some rules, some help to build upon from that.' ''}. 

\subsubsection{Strategies and best practices}
To overcome the challenges, our participants collectively recommended several strategies for facilitating the creation, maintenance, and use of design systems.

First, having a \textbf{dedicated team} that is responsible for developing and maintaining the design system is essential. This practice will help maintain the consistency of the design, facilitate communication, and provide adequate support in maintaining and adopting design systems. The simple fact that a dedicated team is created will actually help the design system approach to be better integrated into the company's culture. For example, P5 described: \textit{``I think that you have to have kind of ambassadors of a design system within a company to actually get that to be continuously used.''}

Many participants also emphasized the importance to follow a \textbf{bottom-up approach} in which components and guidelines in design systems are emerged and elevated from existing digital products in the company. This will allow more flexibility and facilitate the adoption of the design system, in contrast to a top-down approach in which the design system is created out of new design ideas and impose significant changes to existing products. For example, P1 had extensively discussed this issue: \textit{``It's sort of a push and pull thing. Sometimes you get a new piece of user experience that is created for specific products. Then the design system team might go and look at that and say, "Actually, that is quite a useful widget. We're going to make a generic version of that." Then that generic version can get pushed back down into other products so that everyone is using a consistent version. ... So you don't want to say, "All the designers now work in a design system team. And if you need anything in your products that require design then you have to get from the design system."''} 


Participants also emphasized that the design system teams need to be \textbf{proactive in communication} in order to effectively incorporate feedback from users of the design system and facilitate its adoption. For example, P5 described: \textit{``You have to set in place processes that allow teams that are not working on, but using, the design system to feedback into the design system.''} P1 also mentioned: \textit{``The people who feel like they put a page on the Wiki somewhere and therefore everyone in the organization knows what's going on -- that's a complete travesty. ... It doesn't matter how much you're communicating, it's not enough, communicate more. ... Then you will start to approach levels of ease of use and organizational knowledge which are actually where you want to be.''}

Finally, having \textbf{high-quality documentation} was also regarded by participants as an essential element for any design system project to succeed. Overall speaking, good documentation increases development speed, resolves communication challenges, and facilitates the use of design systems. For example, P4, who focuses on the design system in a relatively small organization, mentioned: \textit{``We don't have the same number of users Material Design has, but somehow we need to serve them the same way. If we didn't have the right documentation we would have failed in the first year probably.''} P9 also considered the development experience and adoption when discussing this aspect: \textit{``You have to spend so much time on documentation because it goes all the way to the developer experience.''}

\section{Discussion}
Our study is aimed at understanding the current design system practice and discovering opportunities for tools and methods that can potentially support this practice. Based on the insights gained from our studies, we could provide a living definition of design systems, from the point of view of the practitioners who create and maintain the design systems. Our living definition defines a design system as \textit{\textbf{an all-in-one design and development environment that includes both UI components and guidelines and serves as a common language for unifying design.}} We encourage the research community to continue to refine and improve this definition. Additionally, we have made several important contributions that we will discuss in this section. First, our analysis of issues discussed by open source design system communities revealed that the communities had diverse concerns and focuses, reflected in the various project aspects these issues touched. Second, our interviews with leaders of the design system projects uncovered prominent challenges and successful strategies that need to be considered when creating tools for supporting the design system practice. And finally, our study provided important implications towards the design of techniques and tools for supporting the design system practice.

\subsection{Issues Raised by Design System Communities}
The design system communities in the open source platform seemed to have similar concerns with other open source communities in terms of the nature of their issues. For example, \cite{6698918} found that open source communities, in general, tend to mostly discuss bugs and request new features in issue tracking systems (ITSs). The majority of issues discussed by design system communities are also of these types.

In terms of the project aspects, issues discussed by the design system communities exhibited some interesting characteristics. Particularly, communities tended to focus on discussing the behavior of the UI components in their design systems (in 42.7\% of the analyzed issues) more than their visual design (in 10.2\% of the analyzed issues). One explanation of the focus of discussion on UI behavior is that the visual design (e.g. color scheme and typography) is more bound to the branding of the products that the design systems facilitate, and thus tend to be centrally managed and more stable. The behavior of the UI components, on the other hand, needs to adapt to the user interaction needs of specific products and thus more subject to change and prone to errors. In fact, the communities have focused on several important aspects of UI behaviors such as accessibility and safety. These aspects often require special expertise to address. Elevating their discussion at the design system level would promote the consistent adoption of the best practices of these aspects across all products.

Overall, the design system communities used the issue tracking systems to discuss various aspects of the design system projects. In addition to the main concern of UI components, these issues included topics related to documentation, software development, and community engagement. These characteristics reflected the fact that design systems are composed of various deliverables, including pattern libraries as UI components and design guidelines and style guides as documentation~\citep{Hacq2018, Vendramini2021}. Naturally, design system issue discussions tend to put emphasis on the design aspects of the project. However, during our analysis we noticed that, even for UI components, both design and development aspects were touched in the discussions. 

Because of the nature of design systems as both component libraries and design guidelines~\citep{Hacq2018, Fessenden2021, Vendramini2021}, we have speculated that documentation issues would also be a frequent topic. To our surprise, however, they only comprised about 10\% of the analyzed issues. Possible explanations of this phenomenon could be that (1) documentation aspects were well-maintained internally that limited the issues exposed to the community outside of the organization and/or (2) the design system open source communities were still not mature enough to break through the common pitfalls of open source development (i.e., focus on system-related aspects rather than user-related aspects~\citep{wang2020open}). Future studies need to be conducted to examine these explanations.

\subsection{Concerns of Design System Project Leaders}
Through our interviews with leaders of prominent design system projects, we have revealed their considerations with regard to the concepts, benefits, challenges, and best practices of design systems. While our participants discussed various benefits the design system approach could offer, all these benefits are connected around the aim of the design systems in creating an all-in-one, centralized environment for user interaction design and development. Such an environment promotes consistency and coherence and facilitates designer-developer collaboration. These advantages in turn result in efficiency and reduced cost in both the design and development work, and thus potentially streamline the designer-developer collaboration. These benefits are in fact presented frequently in the literature that include practitioner-oriented websites and blog posts~\citep{Fanguy2019, Fessenden2021} as well as researcher-oriented articles~\citep{Churchill2019}.

Results from this interview study have also extended our findings of the community interests, revealing concerns and challenges from the internal of the organization that manages the design systems. These challenges are often not discussed in the existing gray or white literature of design systems. First, the leaders' challenge of balancing customizability and standardization corresponds to the numerous issues community members discussed on the behaviors of the UI components. In these issues, users of the design system often requested behaviors that emerged from the needs of their specific products, and the design system team had to make difficult decisions about whether to satisfy these needs at the design system level. Second, our interview participants have discussed extensive concentration and centralized control on the documentation of design systems within the organization. This practice partially explains the fact that documentation issues were less discussed by the broader community; i.e., documentation about design systems is typically well-maintained internally. Finally, evolving design systems is considered to be a prominent challenge by our participants; this problem is often aggravated by ineffective communication and collaboration among stakeholders (particularly among designers and developers). Our participants recommended having a bottom-up approach to create and maintain the design systems in order to keep them relevant and useful to the product teams while the concrete products that use the design system evolve themselves. 

\subsection{Needs of Tools to Support the Design System Practice}
The most notable aspect of the design system practice that has emerged from our studies is the intricate balance that this practice has to address between the stable design knowledge within the organization and the ever-changing needs of users and products. On one hand, practitioners have defined design systems as centralized environments that accumulate and preserve the design knowledge that is ready to be used. This knowledge needs to be stable enough to achieve consistency in products and facilitate communication. On the other hand, based on their immediate projects, the community using the design system constantly had diverse needs and requests that create difficult decision points for the design system maintainers. In order to be successful, the design system approach thus needs to be both rigid and flexible, both stable and open to change. This characteristic of design systems creates an intricate challenge for tools and techniques that can support this approach. Our findings indicated several opportunities and considerations for the design of such tools and techniques.

First, the design system is still a new concept for many organizations and it implies a new mindset and process towards software design and development. It is thus important to investigate tools and techniques that flatten the learning curve and support organizations to kick-start the design system practice. For example, tools can provide templates including the core elements of a design system (e.g., UI components, design guidelines, development documentation, etc.) to support the creation process; alternatively, an initial design system can be created automatically from one or more products to be customized by design system maintainers.

Second, maintaining and evolving the design systems are considered a big challenge. The number and the diversity of the issues raised by communities of open source design systems also highlighted this complexity. Our participants recommended a bottom-up approach that focuses on iteratively extracting, combining, and incorporating UI elements in concrete products into the design system. Tools that could streamline the evolution of design systems through this bottom-up approach can thus ease the effort of design system maintainers. Particularly, it would be useful to explore semi-automated tools that detect similar UI elements in the products to be elevated and merged into design system components. In such tools, design system maintainers should have the ability to consolidate different styles from different products in a centralized way to ensure consistency in the design system. Additionally we identified that, comparing to other aspects of design system projects, behaviors of UI components tend to receive more issues and requests, thus might undergo more frequent changes. So tools and techniques that support maintaining and evolving the design systems may prioritize this project aspect.

Third, related to the previous point, design system components should be customizable in order to meet the specific needs of the products. In open source design system projects, these needs are often presented in issues that are \textit{requests for new components} and \textit{suggestions for improvements}; these issues constituted 37.1\% of those we analyzed. Our participants also indicated that the required level of customizability can be different in different components; e.g., a button component may require a lower level of customizability than a calendar component. Moreover, the variability of a certain type of UI element in the products created in an organization can indicate its required level of customizability in the design system. Thus, design systems tools can be created to help determine the level of customizability, as well as the customizable parameters, by analyzing existing products.

Fourth, documentation such as design patterns, design guidelines, and component development guides are integral elements in a design system. Tools that support the creation and maintenance of the documentation along with the design system components would be useful. Automatic techniques can also be explored to generate the documentation while the components are elevated from UI elements. Additionally, we found that documentation issues are not discussed extensively in open source communities although the community feedback is valued by the design system project leaders during our interviews. Thus, tools that encourage consideration and feedback towards documentation, especially the design patterns and guidelines, would be helpful.

Finally, design system is considered as a catalyst for facilitating the collaboration between designers and developers. As a result, any tools designed for supporting design system creation and maintenance should consider the needs and expertise of both designers and developers. For example, for customizing design system components, the tools should allow designers to perform the customization through a graphic UI and at the same time provide the flexibility for developers to directly edit the code; both approaches need to be synchronized and streamlined. Additionally, design system tools could provide a user interface that is consistent with the existing tools that designers and developers are familiar with.

We are currently working on incorporating some of these considerations in a design system tool through a user-centered design approach. So far, we have created several prototypes and received feedback from six design system experts. Our latest prototype is a web application that supports merging styles from multiple websites to customize the design system components. Our preliminary user studies considered that these features are suitable to fit in the workflow of evolving design systems; many of our participants also commented on the ability of a centralized tool in helping address the collaboration gap between designers and developers.

\section{Threats to Validity}
There are several threats to the validity of our study. First, to understand the issues to be addressed by the design system projects, we used the issue tracking systems of open source design systems as a proxy. This approach acknowledges the trend of open sourcing design system projects and the fact that many important projects are already put open source. However, we were not able to include proprietary design systems or community discussions that happened outside of issue tracking systems (e.g. in forums,  conferences, or internal discussion channels). A broader study with other types of design systems and communication platforms would be able to address this limitation. Second, we performed a random sampling of the issue discussion data. While it allows us to have a sample that represents our population as a whole, our analysis results may be biased towards larger projects with more issues. A project-based analysis may be used to address this limitation. Third, in our interviews, we used a developer-oriented platform (i.e., GitHub) as our main recruitment venue. This recruitment strategy has the risk of missing important contributor roles such as designers and managers. However, we found that these roles are often also active on GitHub repositories of design system projects, as guideline/documentation maintainers and issue organizers/discussants. In fact, our participant sample has a balanced distribution among the three roles (i.e., three designers, three developers, and three managers). Finally, our interviews are only focused on the highly experienced contributors and decision-makers of design systems. While they covered diverse roles and have provided important insights into the whole life cycle of design systems, perspectives from inexperienced practitioners and users can extend these insights in particularly challenging directions such as kick-starting and adopting the design system approach. We encourage the research community to explore these areas in the future.
\section{Conclusion}
In this paper, we reported on an exploratory study that focused on understanding and supporting the now widely-discussed but understudied design system practice. Particularly, we identified the aspects in issues that design system projects often deal with, investigated the perception and practice of design system experts, and explored the needs of tools that support the design system creation and maintenance process. Our studies serve as an important step towards understanding this important user interaction design and development practice, as well as synthesizing knowledge to inform tools to better support this practice. Our findings will inspire more studies to investigate the design system and related modern user interaction design and development approaches.

\begin{acknowledgements}
We thank our participants for their time and valuable feedback. This work was partially supported by the Discovery Grant of the Natural Sciences and Engineering Research Council of Canada [RGPIN-2018-04470].
\end{acknowledgements}

\section*{Declarations}
\noindent\textbf{Conflicts of interests/Competing interests:}
There is no conflict of interest or competing interests involved in this project.

\vspace{8pt}
\noindent\textbf{Ethics approval:} All studies that involve human participants in this project were approved by the ethics review board of Polytechnique Montreal.

\bibliographystyle{spbasic}
\bibliography{DesignSystem,OSS}

\end{document}